\documentclass[usenatbib,a4paper]{mn2e}

\usepackage{xspace}
\usepackage{graphicx}
\usepackage{amsmath}
\usepackage{caption}

\newcommand{\vect}[1]{\boldsymbol{#1}}

\def\spose#1{\hbox  to 0pt{#1\hss}}  
\newcommand{\lta}{\mathrel{\spose{\lower 3pt\hbox{$\sim$}}\raise  2.0pt\hbox{$<$}}}
\newcommand{\gta}{\mathrel{\spose{\lower  3pt\hbox{$\sim$}}\raise 2.0pt\hbox{$>$}}}

\newcommand{\be}{\begin{equation}}
\newcommand{\ee}{\end{equation}}

\newcommand{\citepeg}[1]{\citep[e.g.][]{#1}}

\newcommand{\kms}{\ifmmode  \,\mathrm {km}\,s^{-1} \else $\,\mathrm{ km\,s^{-1} } $ \fi }
\newcommand{\kpc}{\ifmmode  {\mathrm{kpc}}  \else ${\mathrm{  kpc}}$ \fi  }  
\newcommand{\pc}{\ifmmode  {\mathrm{ pc}}  \else ${\mathrm{ pc}}$ \fi  }  
\newcommand{\Msun}{\ifmmode {\mathrm{ M_{\odot}}} \else ${\mathrm{ M_{\odot}}}$ \fi} 
\newcommand{\Zsun}{\ifmmode {\mathrm{{Z_{\odot}}}} \else ${\mathrm{ Z_{\odot}}}$ \fi} 
\newcommand{\yr}{\ifmmode yr^{-1} \else $yr^{-1}$ \fi} 
\newcommand{\hMsun}{\ifmmode h^{-1}\,\rm M_{\odot} \else $h^{-1}\,mathrm{\ M_{\odot}}$ \fi}

\def\pr{{\mathrm{ Pr}}}

\def\spose#1{\hbox  to 0pt{#1\hss}}  
\renewcommand{\lta}{\mathrel{\spose{\lower 3pt\hbox{$\sim$}}\raise  2.0pt\hbox{$<$}}}
\renewcommand{\gta}{\mathrel{\spose{\lower  3pt\hbox{$\sim$}}\raise 2.0pt\hbox{$>$}}}

\renewcommand{\be}{\begin{equation}}
\renewcommand{\ee}{\end{equation}}

\newcommand{\bea}{\begin{eqnarray}}
\newcommand{\eea}{\end{eqnarray}}

\def\pr{{\mathrm P}}

\newcommand{\comment}[1]{}
\newcommand{\comments}[1]{}

\newcommand{\apj}{ApJ}
\newcommand{\apjl}{ApJL}
\newcommand{\apjs}{ApJS}
\newcommand{\mnras}{MNRAS}

\newcommand{\aap}{A\&A}

\newcommand{\prd}{Phys. Rev. D}
\newcommand{\nat}{Nature}

\newcommand{\url}[1]{#1}

\def\icg{Institute of Cosmology and Gravitation, University of Portsmouth, Burnaby Rd, Portsmouth, PO1 3FX, UK}

\def\collettemail{\tt thomas.collett@port.ac.uk}

\date{Submitted to MNRAS}

\title[Selection biases in time-delay cosmography]{Observational selection biases in time-delay strong lensing and their impact on cosmography}
    
\author[Collett \& Cunnington, 2016]{%
  Thomas~E.~Collett\thanks{\collettemail}, Steven.~D.~Cunnington
  \\\icg
}

\begin{document}

\pagerange{\pageref{firstpage}--\pageref{lastpage}}\pubyear{2016}

\maketitle 

\label{firstpage}

\begin{abstract}

Inferring cosmological parameters from time-delay strong lenses requires a significant investment of telescope time; it is therefore tempting to focus on the systems with the brightest sources, the highest image multiplicities and the widest image separations. We investigate if this selection bias can influence the properties of the lenses studied and the cosmological parameters that are inferred. Using a population of lenses with ellipsoidal powerlaw density profiles, we build a sample of double and quadruple image systems. Assuming reasonable thresholds on image separation and flux, based on current lens monitoring campaigns, we find that the typical density profile slopes of monitorable lenses are significantly shallower than the input ensemble. From a sample of quadruple image lenses we find that this selection function can introduce a 3.5\% bias on the inferred time-delay distances if the ensemble of deflector properties is used as a prior for a cosmographical analysis. This bias remains at the 2.4\% level when high resolution imaging of the quasar host is used to precisely infer the density profiles of individual lenses. We also investigate if the lines-of-sight for monitorable strong lenses are biased. After adding external convergence, $\kappa$, and shear to our lens population we find that the expectation value for $\kappa$ is increased by 0.004 and 0.009 for doubles and quads respectively. $\kappa$ is degenerate with the value of $H_0$ inferred from time delays; fortunately the shift in $\kappa$ only induces a 0.9 (0.4) percent bias on $H_0$ for quads (doubles). We therefore conclude that whilst the properties of typical quasar lenses and their lines-of-sight do deviate from the global population, the total magnitude of this effect is likely a subdominant effect for current analyses, but has the potential to be a major systematic for samples of $\sim$25 or more lenses.
\end{abstract}


\setcounter{footnote}{1}


\section{Introduction}
\label{sec:intro}

Over the next decade several telescopes will conduct deep, wide sky surveys, with the goal of understanding the Dark Energy that is thought to drive the accelerated expansion of the Universe.  Currently the data is consistent with Dark Energy being the cosmological constant \citep[e.g.][]{planckXVI,boss2015,SVdesWL,betoule,collett2014}, but the large uncertainties on these measurements mean that many Dark Energy models can still fit the data. Making progress requires an increase in measurement precision whilst simultaneously ensuring systematic errors are controlled.

Strong gravitational lensing is one of a small number of probes that can make high precision measurements of cosmological parameters \citep[e.g.][]{refsdal,collett2012,grillo,jee2015}. In strong lens systems where the source is time variable, the multiple images will not vary simultaneously, since the light travels along different path lengths and traverses different parts of the gravitational potential field. The time-delay between the images is primarily a function of the image positions, the mass profile of the lens, and the time delay distance which is a function of the source and lens redshifts and the cosmological parameters. Measurements of the time-delay distance are primarily sensitive to the Hubble constant, $H_0$. Furthermore, time-delay lenses are also highly complementary, when added to other cosmological probes, in determining the dark energy equation of state \citep{linder2011}.

The cross-section for strong lensing and how it changes with the lens parameters has been investigated in previous work \citep[e.g.][]{mandelbaum}, but these results have only focused on the probability for multiple imaging to occur. In the next three paragraphs, we detail the complications of time-delay cosmography, that place selection pressures on the population of lenses that are studied in detail. In this work our goal is to assess how the observational selection effects of telescope limiting magnitude and resolution affect the the probability of {\it detectable} multiple imaging to occur. 

The first difficulty with conducting time-delay science is obtaining precise time-delays; regular and long-term monitoring campaigns are necessary \citep[e.g.][]{bonvin2015}. The required amount of observing time means that mostly small, $\sim$1m, telescopes have been used for lens-monitoring. With these telescopes it is only plausible to obtain time-delays for the brightest and widest separation lenses. The typical image separation required for time-delay estimation by the Cosmograil collaboration is $1.5$ arcseconds, and the typical minimum magnitude is 19 for each image (F. Courbin, Private communication).

The second difficulty for time-delay science is constraining the mass profile of the lens. \citet{suyu2010} showed that high resolution imaging  of a quadruple-image quasar system combined with precise time-delays allow for competitive cosmological constraints, measuring $H_0$ to be $70.6\pm 3.1$ km s$^{-1}$ Mpc$^{-1}$. In \citet{suyu2013} a second quadruple-image lens was used to infer $H_0$ to be $75.2^{+4.4}_{-4.2}$ km s$^{-1}$ Mpc$^{-1}$ assuming a $\Lambda$CDM with a WMAP7 prior \citep{wmap7}. Focusing on quadruple-image systems has the advantage that the time-delays and image positions place significantly more constraints on the lens model than in a double-image system. For a double image system, model assumptions must be made \citepeg{paraficz2009}, or high resolution imaging of the quasar host galaxy must be obtained \citep{suyu2012doubles}. However, focusing on quads has the potential to bias the inference on cosmological parameters, if these systems are a biased subset of the lens population.

The final difficulty in using time-delays to constrain cosmological parameters is overcoming the mass-sheet degeneracy \citep{msd} and the source-position transformation \citep{spt}. Imaging data alone are identically well fit by both a lens with density profile $\kappa(\vect{x})$ and a mass-sheet transformed $\kappa(\vect{x})^{\prime}=(1-\lambda)+\lambda \kappa(\vect{x})$ \footnote{the unknown source must also be scaled}. However the time-delays are proportional to $\lambda$, so the inferred time-delay distance is degenerate with the unknown value of the rescaling $\lambda$.  This mass-sheet can be internal to the lens where the assumed density profile deviates from the true profile by a mass-sheet transformation within the Einstein ring \citep{xu+sluse2015}, or external to the lens where the outskirts of dark matter halos along the line-of-sight act like mass-sheets plus an external shear. The internal mass-sheet can in principle be broken with observations of the stellar kinematics, and whilst inference on the external mass-sheet can be made using observations of galaxies along the line of sight \citep{collett2013, greene}, such measurements are extremely challenging.

\begin{figure*}
  \centering
  \includegraphics[width=0.8\textwidth,clip=True]{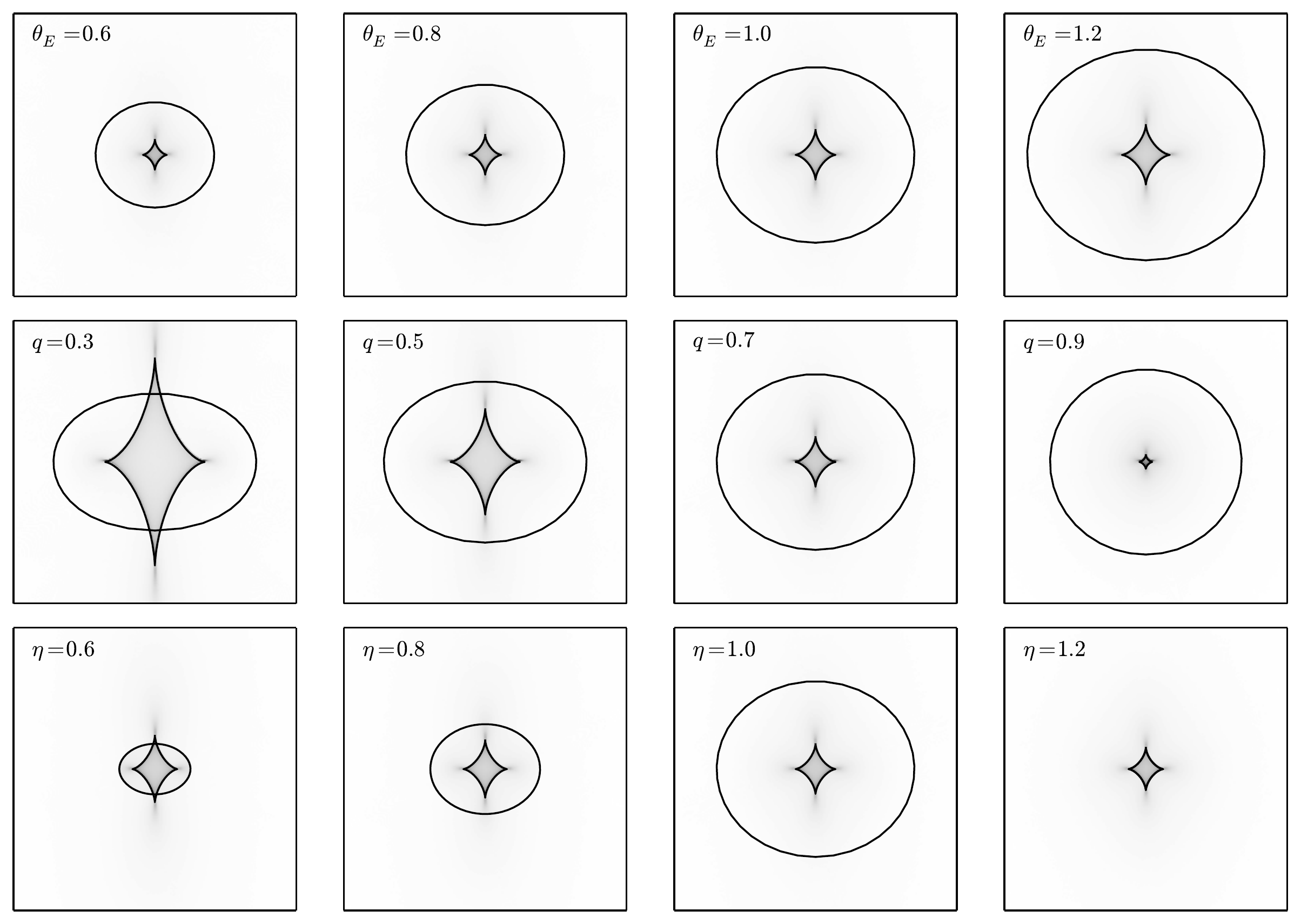}\\
    \caption {The caustics of an elliptical powerlaw density profile lens, as the lens parameters are varied. Sources falling within the central asteroid caustic will form quads, sources outside the asteroid but inside the ellipse will form doubles, sources outside the ellipse are not strongly lensed. The lens model is defined in Equation \label{eq:PL}. The top row shows the effect of increasing Einstein radius, $\theta_E$, the middle row shows increasing the axis-ratio of the lens, $q$, ($q=1$ is a spherical lens) and the bottom row shows increasing powerlaw slope of the density profile, $\eta$ ($\eta=1$ is an isothermal density profile, as $\eta$ increases the central cusp becomes steeper). Unless stated in the top left of each plot, the values of the parameters are set to $\theta_E=1$, $q=0.7$, $\eta=1$. Each plot is to the same scale of 3 units on a side. The bottom right image shows only one caustic, since the deflection angle at $r=0$ is formally infinite for a $\eta>1$, hence double imaging can always occur: in practice images with $r \approx 0$ are highly demagnified.}
    \label{fig:caustics}
\end{figure*}

The primary goal of this work is to answer three questions:
\begin{itemize}
  \item Can focusing scientific resources on bright, wide separation, quadruply imaged quasars introduce a bias on the parameters of the lens?
  \item Can focusing scientific resources on bright, wide separation, quadruply imaged quasars introduce a bias on the properties of the line-of-sight to the lens?
  \item Can these biases introduce significant systematic errors on the cosmological parameters inferred from strong lensing time-delays?
\end{itemize}
A-priori, we expect that the answer to each of the first question will be yes, since the cross-section for a deflector to produce a quadruple image lens depends on the lens parameters, and the properties of the line-of-sight. In Figure \ref{fig:caustics} we show how the caustics evolve as the lens parameters of a power-law ellipsoid change. Only sources falling within the central asteroid caustic form quads and sources outside the asteroid but inside the ellipse form doubles. The rapid evolution of the caustics seen in Figure \ref{fig:caustics} with the flattening of the lens mass and its density slope implies that the flattenings and density slopes of the lens population will be very different to the ensemble of non-lens galaxies.

The paper is organised as follows. In Section \ref{sec:model} we introduce our model for the population of potential deflectors in the Universe. In Section \ref{sec:(Q|L)}, we investigate how changing the lens parameters alters the cross-section for producing detectable lenses. This result is used in Section \ref{sec:results} to infer the distribution of lens parameters given a lensed background source has been observed. In Section \ref{sec:los} we investigate how the line-of-sight might introduce biases. In Section \ref{sec:cosmology} we assess how the results of Sections \ref{sec:results} and \ref{sec:los} affect cosmological parameters inferred from time-delays, and in Section \ref{sec:conclude} we discuss our results and conclude.

\section{A model of the lenses in the Universe}
\label{sec:model}

In order to understand strong lens selection biases, we must first build a model for the population of deflectors and sources within the Universe. For the deflectors, we assume an elliptical power-law density profile; this profile is consistent with the observations of the SLACS lens sample \citep{augerX}. The profile is characterised by three key parameters, the Einstein Radius, $\theta_E$, the axis-ratio or flattening of the lens, $q$, and the powerlaw density profile slope, $\eta$. The reduced surface mass density of the lens is given by:
\be
\label{eq:PL}
\kappa_{\mathrm{lens}}(\vect{x}) = \frac{2-\eta}{2} \left(\frac{\theta_E}{qx_1^2+x_2^2/q}\right)^{\eta}
\ee
where $\vect{x}$ is the position vector relative to the centre of the lens, with $x_1$ aligned with the semi-major axis.

We use the lens population generated by \citet{collett2015}, which draws lens velocity dispersions, $\sigma_V$ from the Sloan Digital Sky Survey (SDSS) velocity dispersion function, as fit by \citet{choi}. We use these velocity dispersions to infer the lens Einstein radius. The lenses are assumed to be uniformly distributed in co-moving volume and the ellipticities of the lenses, $q$, are drawn from P$(q|\sigma_V)$ as fit to SDSS light profiles by \citet{collett2015}. We assume the density slope of the deflectors, $\eta$ is drawn from a Gaussian of width 0.15, centred at 1.08\footnote{an isothermal profile has a slope of 1 in our parameterization, with higher values of $\eta$ corresponding to steeper cusps}, as observed by \citet{augerX}. Since \citet{augerX} finds no evidence that the slope correlates with the mass of the lens, we assume $\eta$ and $\sigma_V$ are uncorrelated.

For the source population we assume the luminosity function of \citet{OM10}. For computational simplicity we assume the sources are on a thin screen at $z_s=1.4$, and neglect the light emitted by the lensing galaxy. Since lens monitoring typically uses telescopes with seeing worse than $\sim$1 arcsecond, we truncate the lens population at a minimum Einstein radius of $0.4$ arcseconds.

\section{Calculating the cross-section for producing double or quadruply imaged quasars}
\label{sec:(Q|L)}

Solving the lens equation for a power-law ellipsoid is computationally expensive. In order to minimise the computational cost of the analysis, we calculate the cross-section for producing doubles and quads using a Monte-Carlo method: For each lens we select 500 source positions $(r_s, \theta_s)$. The unlensed radial coordinate of the source is drawn from a uniform distribution between the projected centre of the lens and 3 times the Einstein radius of the lens, $\theta_e$. This places more sources near the centre of the lens, giving us increased resolution over the small central caustic that produces quads. The angular coordinates are drawn from the uniform distribution in that range 0 to $\pi$. We then assign each source position a weight, $w_s$ such that the total weight is uniform throughout the circle of radius 3$\theta_E$, and proportional to the total area of the disc:
\be
w_{\mathrm{pos}} = 6 \pi \theta_E r_s
\ee

For each source position we then solve the lens equation to find the location at which images form and the magnification of each image. 
The total area of sky that produces a quad is thus given by 
\be
a_Q=\sum_{{\mathrm{quads}}} w_{\mathrm{pos},i}
\ee
where the sum is over all the source positions that produce quads.

In order to assess the detectability of these images, we define a detection magnitude, $m_t$, and a resolution threshold, $R_t$. For doubles we insist that both images are brighter than  $m_t$ and separated by $R_t$, for quads we insist that at least three images are  brighter than $m_t$ and they are separated from each other by at least $R_t$. To avoid the computational cost of including the unlensed source magnitude in our Monte-Carlo, we calculate the maximum unlensed source magnitude that is required to form a detectable double/quad and integrate the quasar luminosity function down to this limit for each source position 
\be
w_{\mathrm{flux}} = \int_{-\infty}^{M_t+\Delta M} \phi(M)dM
\ee
where $M$ is the absolute magnitude of the source, $\phi(M)$ is the quasar luminosity function and $\Delta M$ is given by $\log_{2.5}\mu_i$; $\mu_i$ is the magnification of the second brightest image for a double, and the $i$th brightest resolved image for a quad. We typically chose $i=3$ for quads unless otherwise stated.  $M_t = m_t - \Delta_s$; where the $\Delta_s$ is the distance modulus to the source redshift.

The total weight for producing a quad for each source position is thus given by 
\be
w_Q=\sum_{{\mathrm{quads}}} w_{\mathrm{flux},i}w_{\mathrm{pos},i}.
\ee

We then repeat this process for $2 \times 10^5$ lens models to generate a table of lens properties and weights for each lens to produce quads and doubles.

\section{Results: The cross-section for producing double and quadruply imaged quasars as a function of lens parameters}
\label{sec:results}

Our lens model has three free parameters; $\theta_E, q$ and $\eta$. We are interested in investigating whether $\pr(\theta_E, q,\eta|\mathrm{Q})$ and $\pr(\theta_E, q,\eta|\mathrm{D})$ differ significantly from the input ensemble, which is $\pr(\theta_E, q,\eta)$ for the input deflector population. For each lens parameter, $L$, we can calculate $\pr(L|\mathrm{Q})$ using Bayes theorem to invert the $\pr(\mathrm{Q}|L)$ and $\pr(\mathrm{D}|L)$ inferred in Section \ref{sec:(Q|L)}.

\begin{figure}
  \centering
    \includegraphics[width=\columnwidth,clip=True]{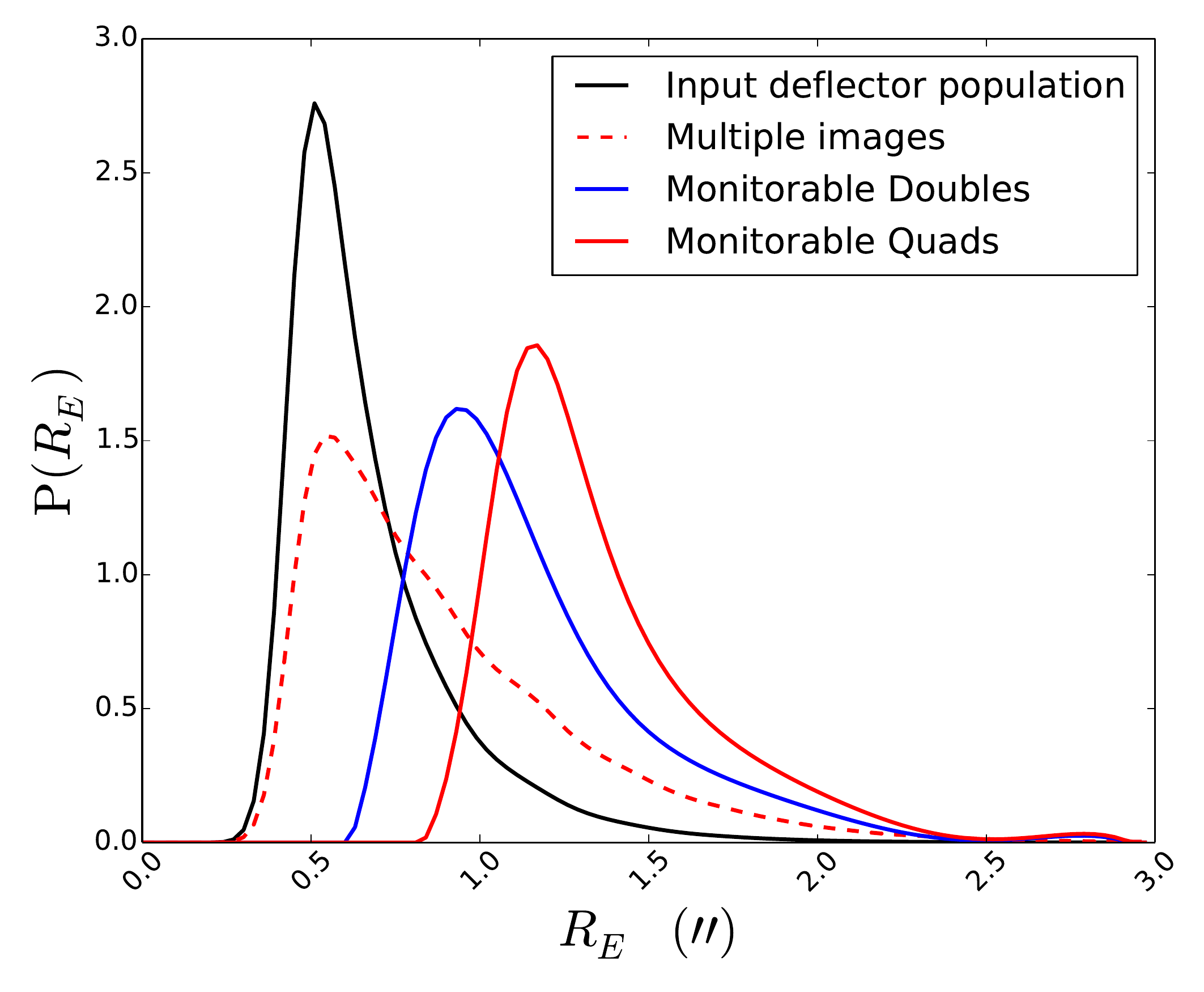}
    \caption {The probability of a lens having a particular Einstein radius. The $\pr(\theta_E)$ distribution for the input ensemble of galaxies is shown in black. $\pr(\theta_E)$, given multiple images are formed is shown as the dashed line. Blue shows $\pr(\theta_E)$ given a double image system forms with both images brighter than $i=19$ and separated by at least 1.5 arcseconds. Red shows $\pr(\theta_E)$ given a quadruple image system forms with three images brighter than $i=19$ and separated by at least 1.5 arcseconds.
    }
    \label{fig:bias_b}
\end{figure}

\begin{figure}
  \centering
    \includegraphics[width=\columnwidth,clip=True]{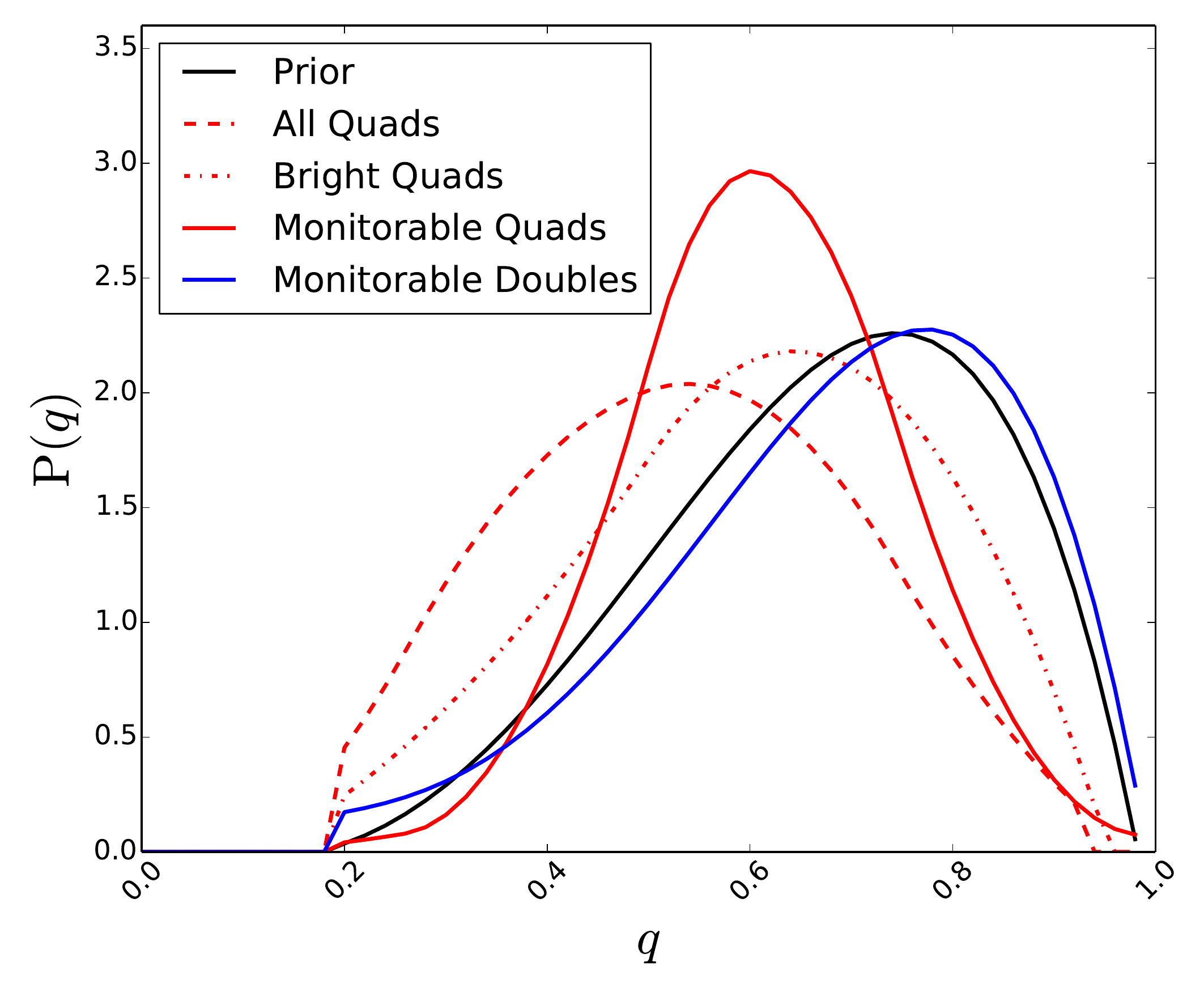}
    \caption {The probability of a lens having a particular flattening $q$. The input flattening distribution for all galaxies is shown in black. The probability of $q$ given four images are formed is shown as the red dashed line. The dot--dashed line is the same but also given that three of the images are brighter than i=19. The red solid line is  $\pr(q)$ given a quad with three images brighter than $i=19$ and separated by at least 1.5 arcseconds. The blue line is $\pr(q)$ given a double with both images brighter than $i=19$ and separated by at least 1.5 arcseconds.}
    \label{fig:bias_q}
\end{figure}

\begin{figure}
  \centering
    \includegraphics[width=\columnwidth,clip=True]{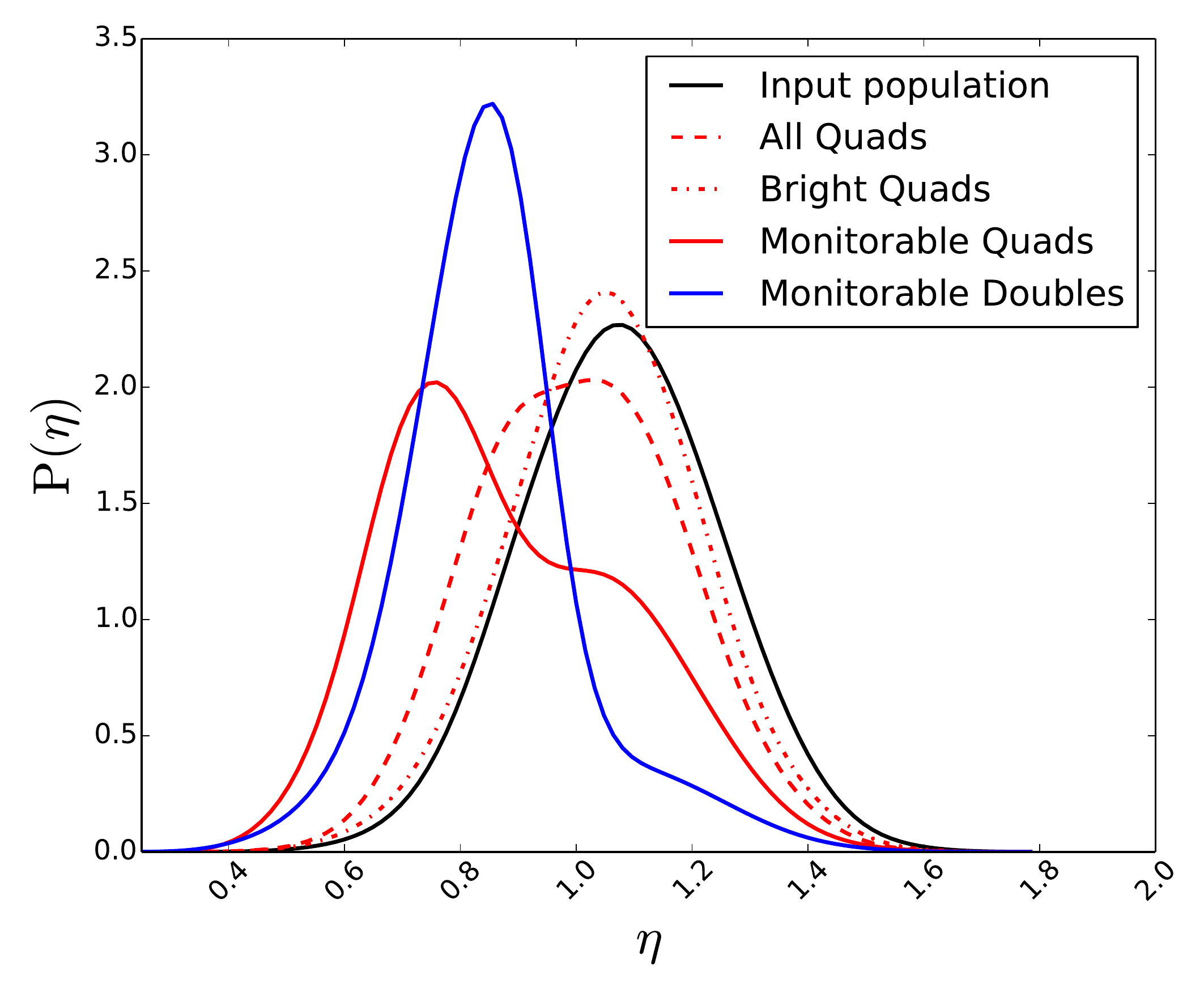}
    \caption {The probability of a lens having a particular powerlaw density slope $\eta$. The input $\eta$ distribution for all galaxies is shown in black. The probability of $\eta$ given four images are formed is shown as the red dashed line. The dot--dashed line is the same but also given that three of the images are brighter than i=19. The red solid line is  $\pr(\eta)$ given a quad with three images brighter than $i=19$ and separated by at least 1.5 arcseconds. The blue line is $\pr(\eta)$ given a double with both images brighter than $i=19$ and separated by at least 1.5 arcseconds.
    }
    \label{fig:bias_n}
\end{figure}

\begin{figure}
  \centering
    \includegraphics[width=\columnwidth,clip=True]{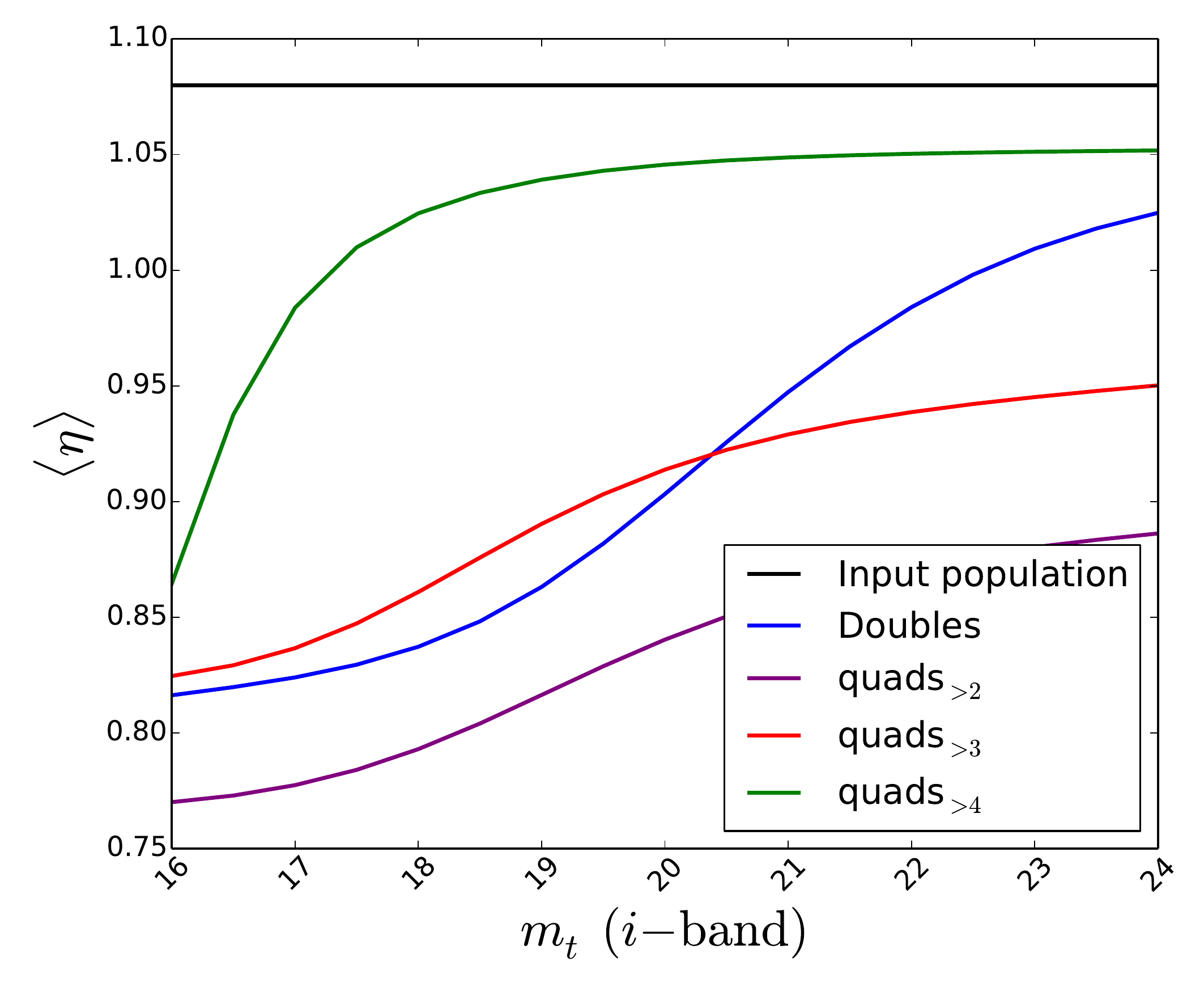}
    \caption {The expectation value for the powerlaw density slope $\langle \eta \rangle$ as a function of limiting magnitude. The resolution threshold is held at 1.5 arcseconds. $\langle \eta \rangle$ for the input galaxy ensemble is shown in black. Blue shows the mean density slope for the population of double image lenses, with both images detected and resolved. The purple, red and green lines are respectively for quads with 2, 3, and 4 images detected and resolved from each other.
    }
    \label{fig:slope_mag}
\end{figure}

\begin{figure}
  \centering
    \includegraphics[width=\columnwidth,clip=True]{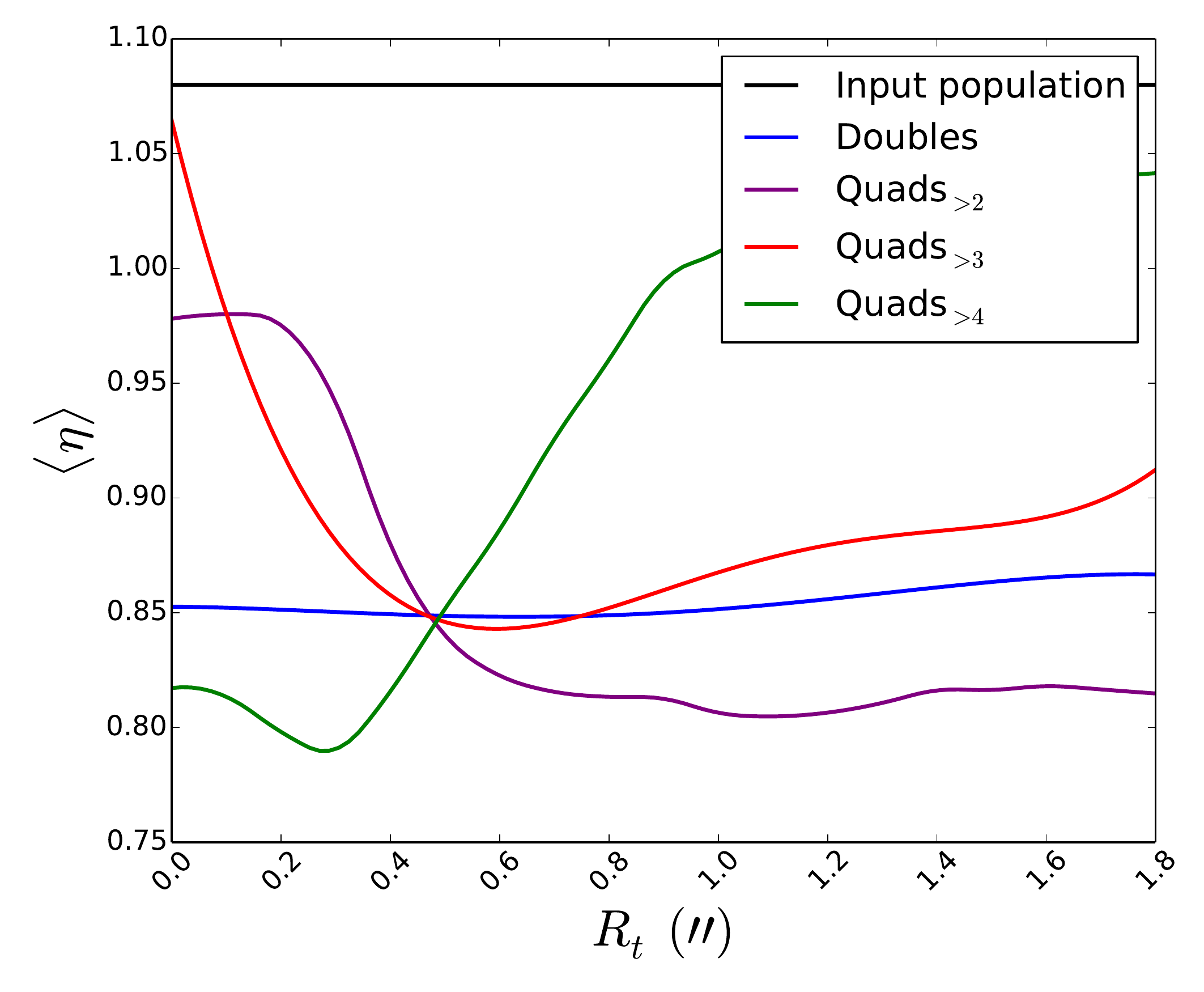}
    \caption {The expectation value for the powerlaw density slope $\langle \eta \rangle$ as a function of limiting image separation. The magnitude threshold is held at $i=19$. $\langle \eta \rangle$ for the input galaxy ensemble is shown in black. Blue shows the mean slope for the population of two image lenses, with both images detected and resolved. The purple, red and green lines are respectively for quads with 2, 3, and 4 images detected and resolved from each other. The lens population is limited to have $\theta_E>0.4$ arcseconds; the low resolution results may not be robust since some lenses with $\theta_E<0.4$ would presumably also be detectable.
    }
    \label{fig:slope_res}
\end{figure}

\subsection{Biases on the Einstein radius of lenses}
Since the linear scale of the caustic structure of a lens is proportional to the Einstein radius, we expect both $\pr(\theta_E|\mathrm{Q})$ and $\pr(\theta_E|\mathrm{D})$ to be proportional to $\pr(\theta_E)\theta_E^2$, however this does not account for the observational requirement that the lensed images be resolved and detectably bright. In Figure \ref{fig:bias_b} we show $\pr(\theta_E|\mathrm{Q})$ and $\pr(\theta_E|\mathrm{D})$. Without including observational effects -- where we have set $R_t=0$ and $w_{\mathrm{flux}}=1$ independent of the magnification -- we find that indeed $\pr(\theta_E|\mathrm{D})$ and $\pr(\theta_E|\mathrm{Q})$ are proportional to $\pr(\theta_E)\theta_E^2$, these are the dashed red line in Figure \ref{fig:bias_b}. There is no change when including a minimum flux threshold on detected images, since the magnification map is self similar under a change in $\theta_E$. Setting the minimum image separation to 1.5 arcsecond however introduces a significant change to the posterior. For a double, (Solid blue in Figure \ref{fig:bias_b}), the image separation is close to 2$\theta_E$: $\pr(\theta_E|\mathrm{D})$ is approximately proportional to $\pr(\theta_E)R_E^2$ for $\theta_E>R_t/2$. Our requirement that quads have 3 resolved images, means that only lenses with an Einstein radius more than approximately $2R_t/3$ are likely to be detectable (Solid red in Figure \ref{fig:bias_b}) .

\subsection{Biases on the ellipticity of strong lenses}
From Figure \ref{fig:caustics} it is clear that more highly flattened lenses have a larger cross-section for producing quadruply imaged lenses than less flattened lenses; indeed it is impossible for a spherical lens to produce a quadruple image system. In Figure \ref{fig:bias_q} we show how $\pr(q)$ changes given an observation of a lens. We find that the expected value of $q$ given the formation of a quad is 0.55; the input population has $\langle q \rangle=0.68$. However the requirement that three images be resolved ($R_t=1.5$ arcseconds) and brighter than $i=19$, increases $\langle q \rangle$ to 0.62 (solid red line in Figure \ref{fig:bias_q}). This increase is due to two effects: firstly changing $q$ affects the magnification and separation of images, and secondly the more massive deflectors in our model are more spherical than less massive deflectors \citep{collett2015}. We find that $\pr(q)$ given formation of a double is comparable to the prior.

\subsection{Biases on the density profile slope of strong lenses}
The cross-sectional area for producing a quadruply imaged quasar is only marginally changed by altering the power-law index of the lens (Figure \ref{fig:caustics}); in Figure \ref{fig:bias_n}, we show that given formation of a quad $\pr(\eta)$ shifts to slightly lower values than for the input ensemble, with $\langle \eta \rangle$ = 1.00. Requiring that three of the images be brighter than $i=19$ negates this shift, with $\langle \eta \rangle$ = 1.04, however once we insist that at least three images be separated by 1.5 arcseconds or more and have $i<19$, we find that $\langle \eta \rangle$ = 0.88. This is significantly lower than mean value for the input deflector population which has $\langle \eta \rangle$ = 1.08. We find the same effect for doubles; requiring both images to have $i<19$ and by separated by at least 1.5 arcseconds gives $\langle \eta \rangle$ = 0.86. This bias towards less cuspy profiles is caused by the fact that more cored profiles tend to produce images with larger separations between bright components. 

In Figure \ref{fig:slope_mag}, we show that the bias on $\eta$ gets smaller as the magnitude limit increases, but does not disappear entirely; with $\langle \eta \rangle = 0.94$ and 1.02 for quads and doubles respectively, with an $i$-band detection limit of 24. We find that changing the resolution threshold makes no significant change to $\langle \eta \rangle$ in the range $0.8''<R_t <1.5''$ (Figure \ref{fig:slope_res}). For quads, the bias is negligible for $R_t < 0.1''$, but this would require a monitoring campaign with either an adaptive-optics assisted telescope or a space telescope which is not currently a feasible proposition. It also neglects the population of lenses with $\theta_E<0.4''$ which are not included in this work. One lens property that has a significant effect on $\langle \eta \rangle$, is the number of bright resolved images that need to be detected. For systems that have two or more images brighter than $i=19$ and resolved by $1.5''$ or more $\langle \eta \rangle = 0.81$, however if we insist that all four images are bright and resolved, we find $\langle \eta \rangle = 1.04$, which is only slightly less than the input although the deviation gets stronger as either of the magnitude and resolution thresholds decrease. 


\section{The effect of line-of-sight lensing}
\label{sec:los}

\begin{figure*}
  \centering
  \includegraphics[width=0.8\textwidth,clip=True]{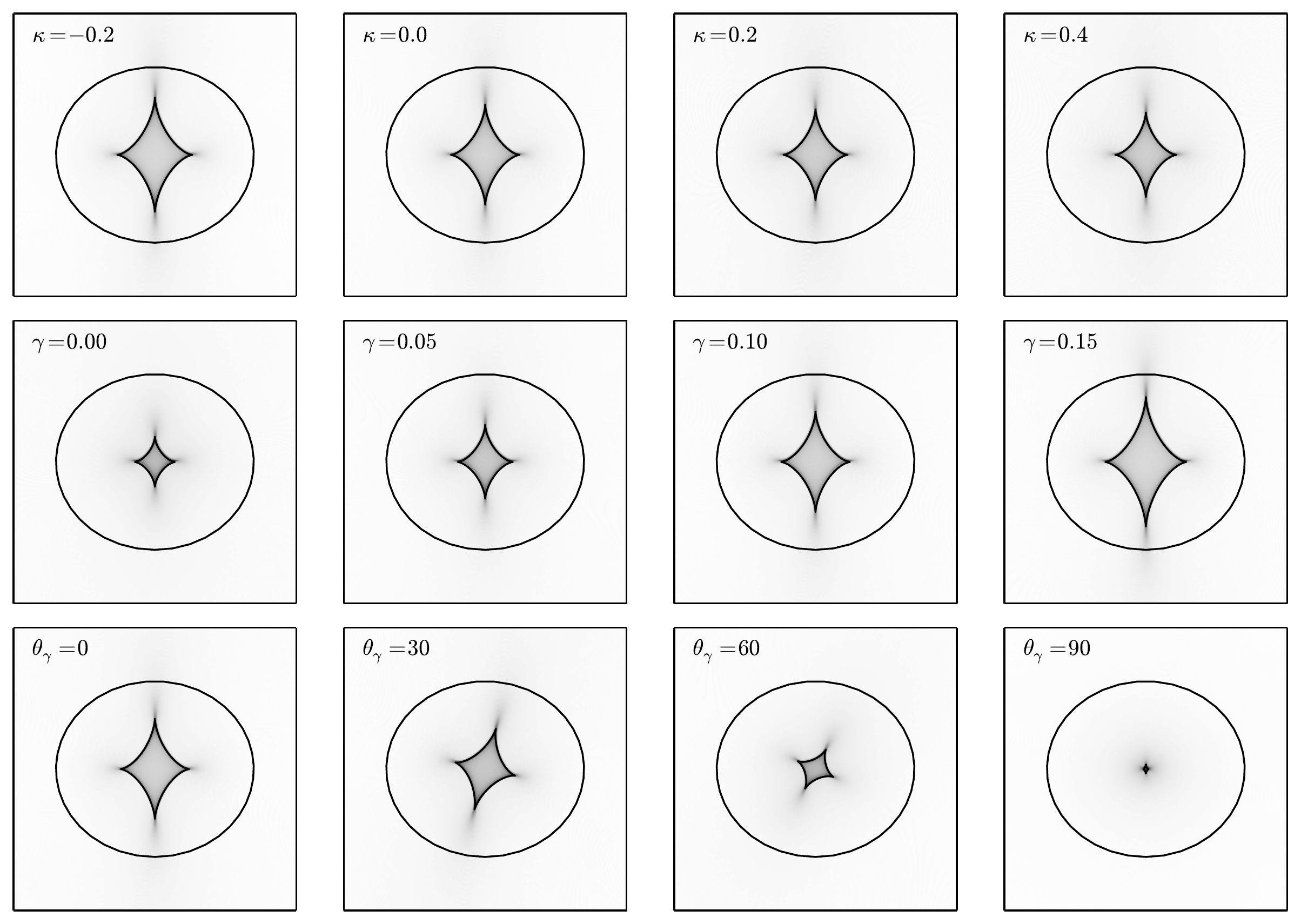}\\
    \caption {The caustics of a powerlaw elliptical lens plus an external convergence and external shear. The top row shows increasing external convergence, $\kappa$, the middle row shows increasing shear, $\gamma$, with the shear aligned with the lens axis ratio. The bottom row shows increasing the angle between the major axis of the lens and the external shear, $\theta_\gamma$. Unless stated in the top left of each plot, the values of the lens parameters are set to $\theta_E=1$, $q=0.7$, $\eta=1$, $\kappa=0$, $\gamma=0.1$, $\theta_\gamma=0$. Each plot is to the same scale of 3 units on a side.}
    \label{fig:caustics_external}
\end{figure*}

\begin{figure}
  \centering
    \includegraphics[width=\columnwidth,clip=True]{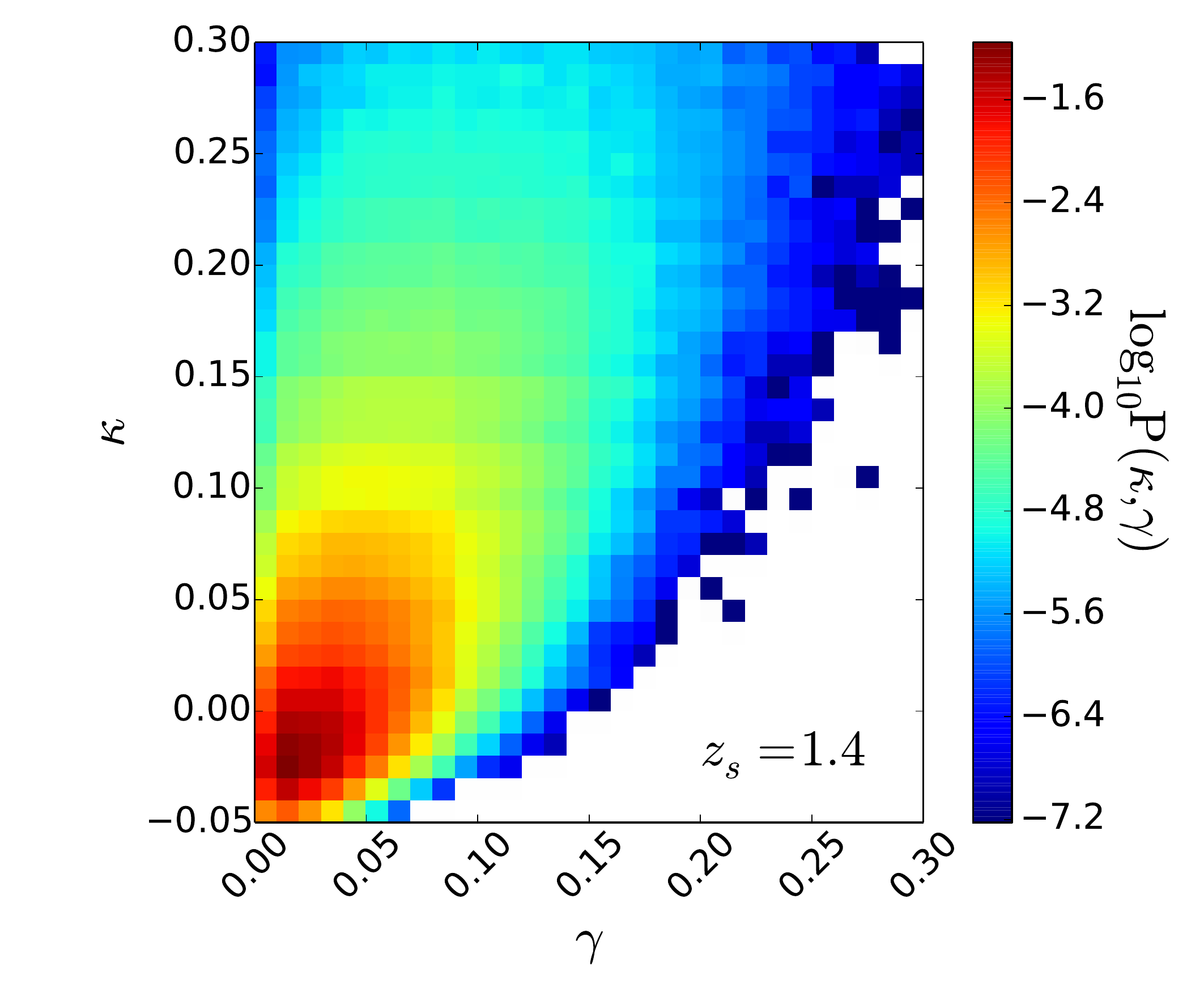}
    \caption {The joint probability distribution of a line of sight having a particular pair of $\kappa$ and $\gamma$ for Millennium Simulation lines-of-sight acting on a source at $z_s=1.4$. The colour scale is logarithmic. White regions had no lines-of-sight with these $\kappa$, $\gamma$ pairs in a 16 square degree patch of sky sampled on a square grid spaced by 3.5 arcseconds.
    }
    \label{fig:Pkg}
\end{figure}

Whilst in Section \ref{sec:results} we showed that the properties of the lensing galaxy alter the probabilities of forming a double or quad, this was under the assumption that the lens is the only mass in an otherwise homogeneous Universe. In reality the Universe is clumpy, and mass along the line-of-sight perturbs the path of light rays through the Universe. If not adequately accounted for, these perturbations can introduce biases on the inferred time-delay distance at the tens of percent level \citep{hilbert2009}; indeed \citet{suyu2014} found that the line-of-sight effects dominate the error budget in their analyses of B1608+656 and RXJ1131-1231.

Large scale structures are well approximated by a quadrupole lens, characterised only by two components; an external shear, $\gamma$, and an external convergence, $\kappa$ \citepeg{hilbert2009}. The correct description of line-of-sight lensing is more complicated than the quadrupole prescription \citep{mccully}, but we leave any investigation of higher order terms to future work. In Figure \ref{fig:caustics_external} we show how the caustic structure of the powerlaw lens changes in the presence of an external shear and convergence. Changing the external convergence produces only small changes to the caustics (although it does affect the locations at which lensed images form). The external shear has a significant effect on the asteroidal caustic that corresponds to formation of quads. When the shear and lens ellipticity vector are aligned, the asteroid grows with increasing shear. When the shear and lens are perpendicular, the asteroid is typically smaller than the no shear case, unless the lens is close to spherical or the shear is very large.

\begin{figure}
  \centering
    \includegraphics[width=\columnwidth,clip=True]{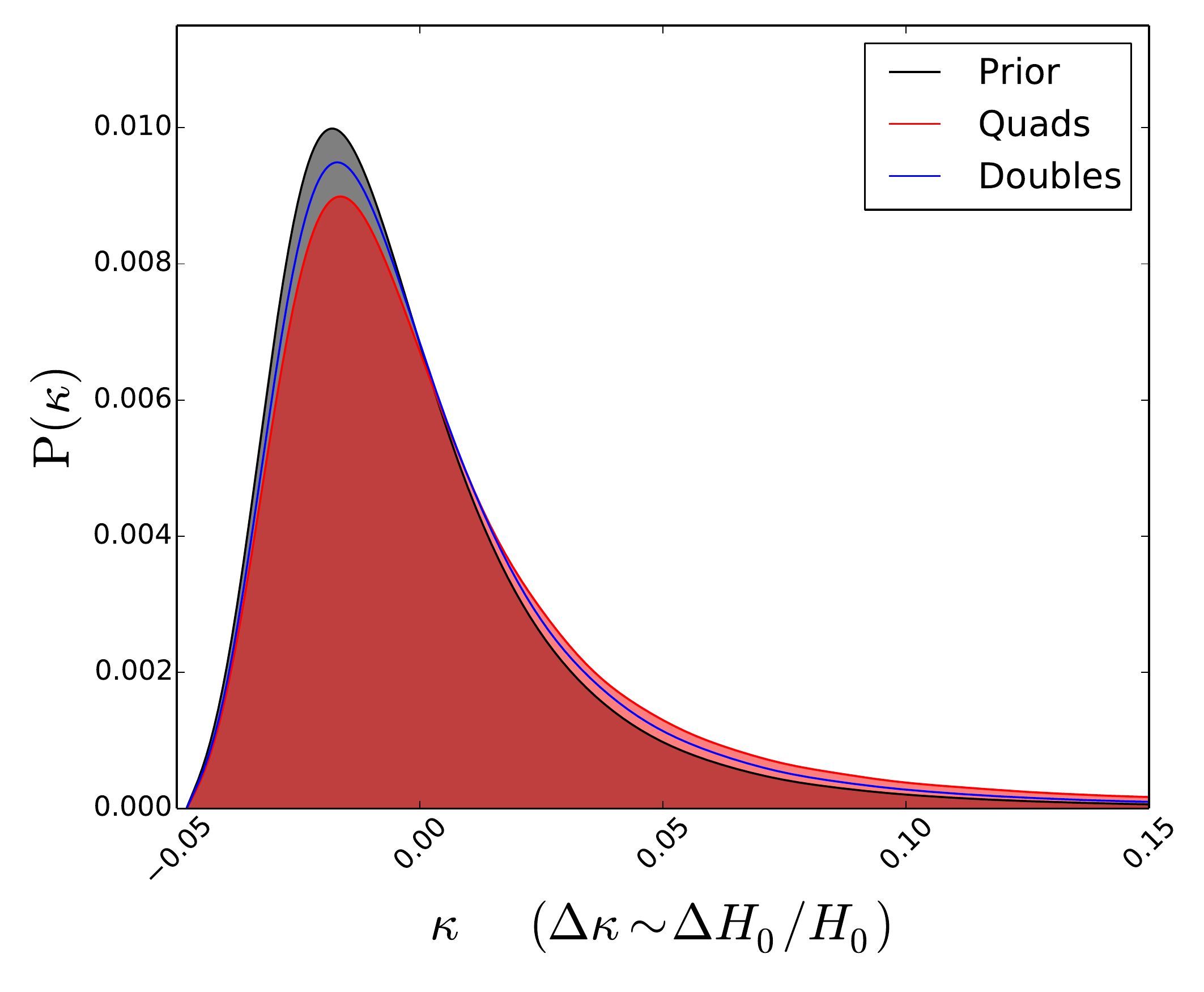}
    \caption {The probability of a lens having a particular external convergence $\kappa$. $\pr(\kappa)$, as derived from the Millennium Simulation, is black. Red is $\pr(\kappa)$ given a quad with three images brighter than $i=19$ and separated by at least 1.5 arcseconds. The blue line is $\pr(\kappa)$ given a double with both images brighter than $i=19$ and separated by at least 1.5 arcseconds.
    }
    \label{fig:kappa}
\end{figure}

\comments{
\begin{figure*}
  \centering
    \includegraphics[width=0.7\textwidth,clip=True]{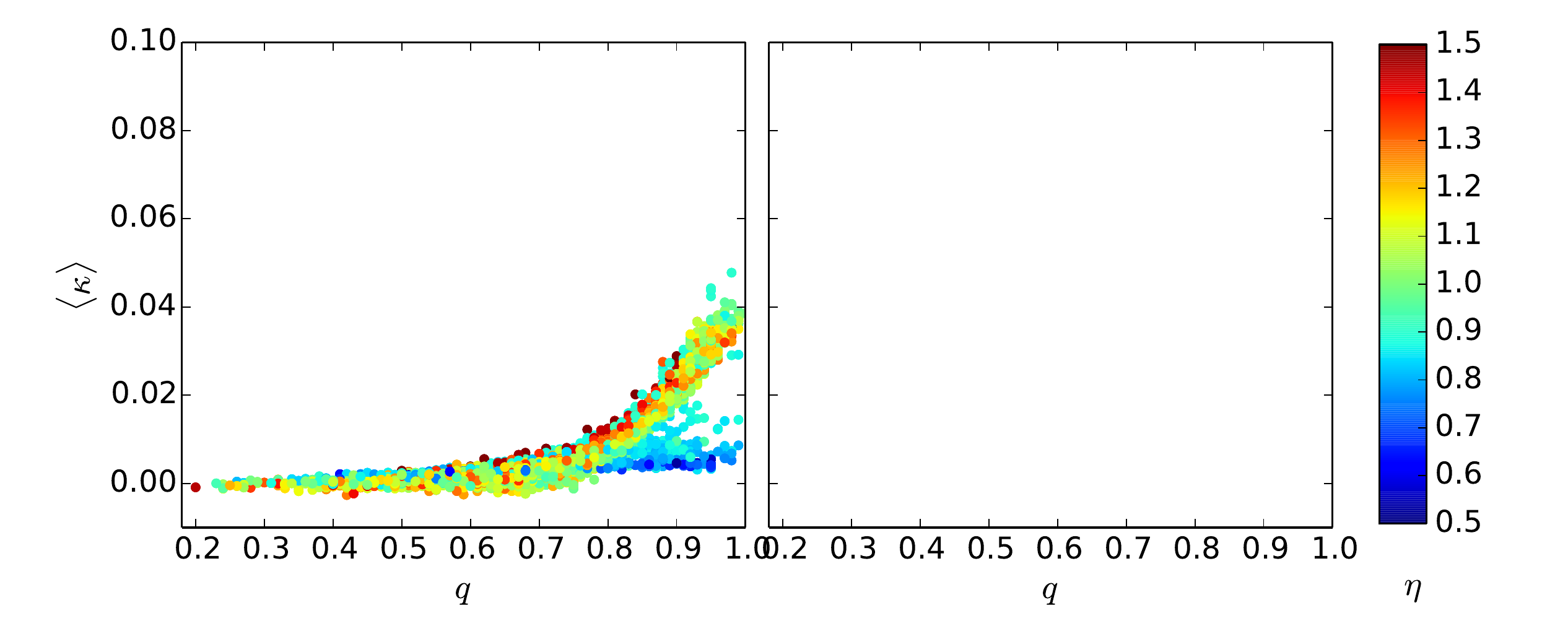}
    \caption {The expected value of external convergence $\kappa$, given formation of a quad as a function of the flattening of the lens $q$. Points are coloured by the powerlaw slope of the lens. The left panel, shows $\langle \kappa \rangle$ for each lens, given that a point source has been quadruply imaged. The right panel shows $\langle \kappa \rangle$ for each lens, given that a quasar has been quadruply imaged with at least 3 images brighter than $i=19$ and separated by at least 1.5 arcseconds. The scatter in the left panel is since $\langle \kappa \rangle$ has been averaged over ten orientations of the lens with respect to the shear. The scatter in the right panel is dominated by the effect of the Einstein radius on the requirement that three bright images be resolved.
    }
    \label{fig:sphericity}
\end{figure*}
}

Ray tracing through cosmological simulations has shown that $\kappa$ and $\gamma$ are correlated, such that high shear lines-of-sight tend to also be over-dense (high $\kappa$). In \citet{suyu2010}, ray-tracing was performed through the Millennium Simulation \citep{millennium} up to a source at redshift 1.4, we use these results to build $\pr(\kappa,\gamma)$, which is shown in Figure \ref{fig:Pkg}. The external shear can be fit for in the lens model, but the presence of an external convergence is undetectable from the observed images due to the mass-sheet degeneracy \citep{msd}. A prior on $\kappa$ must therefore be adopted to infer cosmological parameters from time-delays. 

Taking the distributions of $\kappa$, $\gamma$ and $\theta_\gamma$ from the Millennium Simulation and assuming they do not correlate with the lens properties, we can repeat the analysis of Sections \ref{sec:model} to ask if the prior on $\kappa$ should be changed once we have selected bright wide-separation quads to do our analysis. Due to the computational cost of the modelling we restrict our analysis to 50,000 lenses, and compress $\kappa$ and $\gamma$ onto a single additional potential on the lens plane. In Figure \ref{fig:kappa}, we show that there is negligible bias in $\pr(\kappa|\mathrm{Q})$ or $\pr(\kappa|\mathrm{D})$, assuming $m_t=19$ and $R_t=1.5''$. $\langle \kappa \rangle$ is +0.004 and +0.009 for doubles and quads respectively. The lack of a large bias is due to two reasons; firstly changes in $\kappa$ produce only small perturbations of the image positions, so magnifications and image separations are broadly unchanged - very few lenses change from unresolved to resolved when $\kappa$ increases by 0.1. Secondly, our assumption that the $\theta_\gamma$ is randomly oriented with respect to the lens, means that the external shear is almost as likely to decrease the size of the quad caustic as it is to create an increased cross-sectional area for quadruple imaging. Our results do not change significantly if we increase the detection threshold to $i=21$ and the resolution to $1''$, implying that this result will also hold for future time-delay lens samples.

For lenses that are close to $q=1$ there is a larger bias towards higher $\gamma$ for quads which in turn gives a bias to higher $\kappa$; isothermal lenses with $q=0.9$ that form quads have $\langle \kappa \rangle=0.02$, this increases to 0.03 for lenses with $q=0.95$. However as we found in Section \ref{sec:results} that most quads have lenses with $q \approx 0.6$, the $\kappa$ bias is smaller for most of the population. For almost spherical lenses with $\eta<1$ at fixed external shear a positive $\kappa$ slightly decreases the area enclosed by the quad caustic. The bias towards higher $\gamma$ is still larger than this effect, but the net $\kappa$ bias is smaller for lenses with  $\eta<1$ than for lenses with $\eta>1$.


\section{The impact on cosmological parameters inferred from time-delays.}
\label{sec:cosmology}

In the previous sections we have shown that the properties of the lens population deviate significantly from the properties of the input deflector ensemble. Similarly the lines-of-sight of observable quasar lenses, are not randomly drawn from the distribution of lines-of-sight in the Universe. These results are integral in translating constraints on quasar lens populations into an understanding of galaxy evolution, but the impact of these selection functions on cosmological parameter estimation from time-delay lenses is less obvious.

\subsection{Cosmological biases from the External Convergence}
Because of the mass-sheet-degeneracy $\kappa$ cannot be inferred from lensing data alone.
However, for a $\kappa$ on the lens plane the fractional bias on the time-delay distance is the absolute bias on $\kappa$. An ensemble analysis that assumes $\langle \kappa \rangle=0$ for a sample of doubles and quads will therefore systematically overestimate $H_0$ by 0.4 percent and 0.9 percent respectively, if the lenses are selected with $m_t=19$ and $R_t=1.5''$.

\subsection{Emulating an ensemble analysis with many time delay lenses}

Unlike $\kappa$, biases on the other parameters do not map trivially onto bias on cosmological parameters. A full investigation of how observational selection biases in time-delay strong lenses can propagate into systematic errors on cosmological parameter estimates will depend on the specifics of the lens sample, the data quality and the analysis method used, however we can estimate the likely magnitude of the effect by performing a mock analysis on a sample of lenses (and lines-of-sight) drawn from the selection function derived in the previous sections, but using the parameters of the input deflector ensemble as the priors for the mock cosmological analysis.

For the mock analysis we draw 100 quads assuming $m_t=19$ and $R_t=1.5''$. We model each lens assuming that the time-delays are measured with 1 day precision, and lensed image positions can be measured to $0.025''$. We assume Gaussian errors for the position and time-delays. We do not incorporate any information coming from the relative fluxes of the images, since these are often affected by milli- and microlensing \citep{witt1995}. For each lens we describe the mass with 8 free parameters: the Einstein radius, the powerlaw profile slope, two parameters for the lens centroid, the  flattening of the lens and orientation, and the magnitude and angle of the external shear, there are a further two parameters per system for the unlensed source position. The last parameter is $H_0$, which is a global parameter across all the lenses. Since the external convergence cannot be inferred from lensing alone, we fix $\kappa$ in our mock analysis to the true value for each lens. We assume a flat $\Lambda$CDM cosmology with $\Omega_M=0.3$. We model each system individually, to give a P($H_0$) for each system, the product of which gives the final inference on $H_0$. For our realisation of 100 quadruple image lenses we find $H_0=72.6^{+1.9}_{-2.5}$~km~s$^{-1}$~Mpc$^{-1}$. The input was $H_0=70$~km~s$^{-1}$~Mpc$^{-1}$. Whilst a bias exists at the 3.5 percent level, for this sample, the lens models are only weakly constrained by the image positions and time-delays - only for samples with 100 or more quads is the bias on $H_0$ comparable to the uncertainties.

\comments{We have not investigated the effect of ensemble analyses of doubles, since an 11 parameter model for double image systems is under-constrained using only the positions and time delays. The results of such an analysis would therefore depend entirely on the priors. \citet{wucknitz2002} showed that at fixed external shear, the powerlaw index of the lens and the Hubble parameter are degenerate, with $D_{\Delta t} \propto 2/\eta -1$. }

\subsection{Cosmology with high precision analyses of individual time delay lenses}

An alternative approach to make precise cosmological inference with time-delays is to focus on a small number of systems, but with much improved data quality. This approach has been adopted by \citet{suyu2013} to make a 5 percent inference on $H_0$ with just two lenses. By reconstructing high-resolution imaging of an extended source, the constraints on the lens model are greatly improved. Only a small region of the parameter space can reproduce the quasar positions, time-delays and the arcs with an astrophysically plausible source. Assuming uniform priors \citet{suyu2013} constrains the density slope, $\eta$ of RXJ1131 with precision of 0.05; flattening, $q$, and external shear, $\gamma$, are measured with with 0.007 and 0.006 precision. 

\begin{figure*}
  \centering
    \includegraphics[width=\textwidth,clip=True]{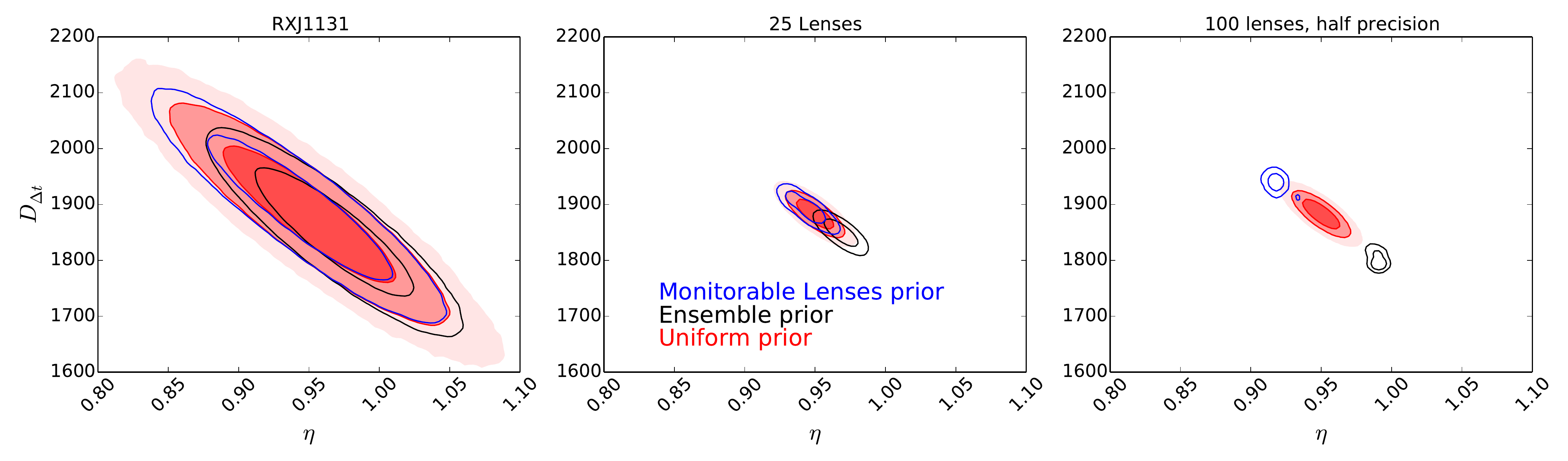}
    \caption {Left Panel: Constraints on the time delay distance of RXJ1131, assuming different priors on the density profile slope $\eta$. The red constraints are the results of \citet{suyu2013} which assume a Uniform prior on $\eta$. Blue contours illustrate the constraints assuming the prior for monitorable quads derived in Section \ref{sec:results}, and black assumes the same prior as that used to generate the ensemble of deflectors in  Section \ref{sec:model}. Middle Panel:  Constraints on the time delay distance made with 25 independent lenses, assuming the measurements for each lens are identical to those from \citet{suyu2013}. Right Panel: Same as middle panel, but for 100 lenses and the precision of the inference on both $\eta$ and $D_{\Delta t}$ are twice that measured for RXJ1131.
    }
    \label{fig:1131}

\end{figure*}

With uniform priors \citet{suyu2013} infers the time-delay distance for RXJ1131 to be $1883^{+89}_{-85}$ Mpc \footnote{neglecting the external convergence from mass along the line-of-sight}. Resampling their results with a prior given by the parameters of the ensemble of deflectors assumed in Section \ref{sec:model} gives a time-delay distance of $1850^{+80}_{-80}$ Mpc. Using the selection function for quads with 3 detectable images separated by 1 arcsecond or more derived in Section \ref{sec:results} we find $1895^{+91}_{-85}$ Mpc. Figure \ref{fig:1131} shows the impact of the different choice priors on the $\eta$-$D_{\Delta t}$ constraints.

The requirement for detectable arcs and the observational cost of deep high resolution imaging will introduce further selection effects on the lens population that are suitable for these analyses. But wherever the lens parameters can be inferred directly from the data with higher precision than the width of the selection function the selection bias will be subdominant. 

When combining many lenses, the statistical errors decrease but the width of the selection function remains the same. In Figure \ref{fig:1131} we show the impact of the selection function on the time delay distance inferred with a sample of 25 lenses like 1131, and 100 lenses like 1131 but with half precision measurements on $\eta$ and $D_{\Delta t}$ for each lens. For the 100 lenses measured at half precision, the results for the three priors differ by significantly more than the precision of the inference. For a real sample of lenses, the uncertainties will vary from system to system and the different values of lens parameters will give different parameter covariances \citep{suyu2012doubles}, but Figure \ref{fig:1131} illustrates how important the choice of priors is likely to be for precision analyses of moderately large lens samples.

\section{Conclusion}
\label{sec:conclude}

In this paper we have investigated how the properties of lens galaxies and the properties of their lines-of-sight affect the probability that they produce bright, large-separation quadruply imaged quasars. We then inverted this probability, to investigate the probability of a lens having specific parameter values, given a bright, large-separation quadruply imaged quasar has been observed. Since current time-delay monitoring surveys such as Cosmograil \citep{cosmograilI} are limited to quasars with images brighter than 19 and separated by at least 1.5'' (F. Courbin, private communication) we focus primarily on quads where at least three images satisfy this criteria. 

We found that quad lenses are likely to be more flattened than the general population, with a median flattening of 0.6. We also found that the power-law slope of monitorable double and quad lenses are significantly shallower than for the input deflector population. \citet{wucknitz2002} showed that at fixed external shear, the powerlaw index of the lens and the Hubble parameter are degenerate, with $D_{\Delta t} \propto 2/\eta -1$. Under the assumption of a powerlaw-ellipsoid lens, the powerlaw index for quads can be inferred from the image positions and time-delays \citep{witt2000} or high resolution imaging of the lensed quasar host \citep{suyu2010}. For doubles, only observations of the lensed quasar host at high resolution can be used to infer $\eta$ \citep{suyu2012doubles}. For a large sample of lenses, such as the 7000 expected in LSST \citep{OM10}, the ensemble of doubles can be used to infer the Hubble constant without high resolution imaging, if a prior on the profile if a prior on $\eta$ is assumed. \citet{oguri2007} used a small sample of quasars, and assumed $\eta=1 \pm 0.15$ for the population to infer $h=0.7 \pm 0.06$. \citet{oguri2007} claim that their value of $h$ is proportional to $2-\eta$; since we find that the detectable doubles have $\langle\eta\rangle=0.86$ this would potentially imply the $H_0$ measurement is biased at the fifteen percent level.

We investigated the posterior for external convergence given a lensed quasar has been observed. The external convergence cannot be inferred from lensing observations alone, and is degenerate with the inferred value of $h$. Using the correct $\pr(\kappa)$ is key to making accurate inference on cosmological parameters. We drew a sample of 50,000 lenses and investigated $\pr(\mathrm{Q}|\kappa,\gamma)$ and $\pr(\mathrm{D}|\kappa,\gamma)$ for each lens. We used ${\kappa, \gamma}$ pairs drawn from the Millennium Simulation. Under the assumption that the $\kappa$, $\gamma$ and $\theta_\gamma$ are independent of the lens parameters, we found a negligible bias; the effect on $h$ is at the $\sim$0.9 per cent level. This assumption may not be valid, since the lens properties are presumably correlated with local structures that contribute to shear and convergence \citep[e.g.][]{IA}. However, this work shows that observation of a multiply-imaged quasar does not significantly bias the part of the line-of-sight that is uncorrelated with the lens. However, we compressed the line-of-sight effect into a single $\kappa$ and $\gamma$ acting on the lens plane; we did not investigate multiple-plane deflections and the non-linearities of multi-plane lensing may amplify the significance of a small line-of-sight bias when inferring cosmological parameters from time-delay lenses.\citep{mccully}

Our results should be interpreted in the light of the assumptions we have made. We assumed a population of elliptical powerlaw deflectors, which is consistent with observations, but is a simplification of the dark and baryonic matter distributions in real galaxies and their substructures. Additionally the input prior on the lens properties is based on the results of a galaxy-galaxy lensing survey \citep{augerX}, which has its own selection function that we have not calibrated for. Our sources are all assumed to lie on a single redshift plane and follow a specific luminosity function, although we do not expect our results to change significantly with source redshift. We have also assumed a simplistic selection function that is a step function based on the number of bright, resolved images. In light of these assumptions, our results cannot be directly applied to any existing lens sample, but provide a robust qualitative understanding of the important observational selection effects.

\begin{itemize}
  \item Can focusing scientific resources on bright, wide separation, quadruply imaged quasars introduce a bias on the parameters of the lens?
\end{itemize}
Yes, quadruple image lenses are significantly more elliptical than the prior $q=0.62$ compared to $q=0.68$. Lensed quasars for which measuring time-delays is possible have significantly shallower profiles than the prior. Quads with at least three images brighter than $i=19$ and separated by 1.5 arcseconds have $\langle\eta\rangle=0.88$. For doubles we find  $\langle\eta\rangle=0.86$. The bias decreases as the magnitude threshold increases. This affect is ameliorated for symmetric quads where all four images are bright and resolved ($\langle\eta\rangle=1.04$), but persists for doubly imaged quasars. 
\begin{itemize}
  \item Can focusing scientific resources on bright, wide separation, quadruply imaged quasars introduce a bias on the properties of the line-of-sight to the lens?
\end{itemize}
Yes but the effect is small, at least not for the part of the line-of-sight that is uncorrelated with the lens parameters. $\langle\kappa\rangle$ shifts by only 0.009.
\begin{itemize}
  \item Can these biases introduce significant systematic errors on the cosmological parameters inferred from strong lensing time-delays?
\end{itemize}
This depends on the data and the analysis. For analyses relying on image positions and time-delays alone the bias on $H_0$ is 3.5\% for quads. For analyses that use high-resolution imaging of the the lensed AGN host galaxy the imaging data overwhelms the small shift in prior. The bias on $H_0$ is likely at the 0.6\% level for the measurement of \citet{suyu2013} using RXJ1131; but the bias becomes more significant if the density profiles of individual lenses are less well constrained by the data, or if multiple lenses are combined. The bias may be larger in current datasets if degeneracies such as the source-position transformation \citep{spt} mean that uncertainties of current measurements of $\eta$ are underestimated \citep{xu+sluse2015}. \citet{meng2015} find that even with faint doubles where total magnitude of the arcs is $\sim$23, Euclid will be able to constrain the slope to 0.034 precision. Since Euclid will cover most of the extra galactic sky, and the width of the selection function we derive for monitorable quads and doubles are 0.2 and 0.15 respectively, it seems that ---unless there are additional systematics in the modelling of Euclid lenses--- the selection function for the density slope should not bias future cosmographic efforts with samples smaller than $\sim$25 time-delay lenses. However, results relying on the combination of many imprecise measurements are particularly sensitive to the sample selection function. 

Our results suggest that the external convergence to the lens will not significantly bias cosmological parameters for doubles or quads. We have not quantified the level of bias on cosmological parameters caused by the part of the line-of-sight that is correlated with the presence of the lens galaxy. Shifts of 0.009 on $\kappa$ give shifts of the same fractional size on $h$. Sub percent precision measurements of $h$ will not be achieved with time-delays in the near future, however systematics of this size may need to be accounted for when combining many cosmological probes in the high precision era of the 2020s.

It is encouraging that lenses do not show a large bias in the physically uncorrelated line-of-sight. Whilst the bias on the density slope can potentially induce large biases on cosmological parameters the fact that this can be overcome with high-resolution imaging of a lensed host, means that this is not likely to be a significant problem for precision analyses of time-delay lenses. This result however serves as a warning that without high-resolution imaging any time-delay cosmography project must think extremely carefully about the prior on $\pr(\eta)$ given the lens selection function. Observing a strongly lensed point source does indeed bias the lens parameters away from those of the general galaxy population, but at most this should only impact the cosmological parameters derived from time-delays at the 2 percent level.


\section*{Acknowledgements}

We thank the referee for their comments and criticisms that have significantly improved this work.
We are grateful to David Bacon, Bob Nichol, Dominique Sluse, Kenneth Wong and Phil Marshall for insightful comments on the original manuscript. We thank Tommaso Treu and Paul Schechter for helpful discussions.

SDC acknowledges funding from the SEPNet summer placement scheme and is grateful to the ICG for hosting the placement.
This project made use of the {\sc LensPop} code which is freely available from \url{github.com/tcollett/LensPop}.

Numerical computations were done on the Sciama High Performance Compute (HPC) cluster which is supported by the ICG, SEPNet and the University of Portsmouth.





\label{lastpage}
\bsp

\end{document}